\documentclass[conference,a4paper]{IEEEtran}

\usepackage{amsmath}
\usepackage{stackengine}
\def\delequal{\mathrel{\ensurestackMath{\stackon[1pt]{=}{\scriptstyle\Delta}}}}
\usepackage{bm}
\usepackage{array}
\usepackage{tabu}

\usepackage{hyperref} 
\usepackage{graphics} 
\usepackage{algorithmic} 
\usepackage{bm}
\usepackage{graphicx}
\usepackage{subfigure}
\usepackage{subfloat}
\usepackage{amsmath}
\usepackage{caption}
\usepackage{amsthm,amssymb,amsmath}
\usepackage{lmodern}
\usepackage{subfigure}
\DeclareMathAlphabet{\mathpzc}{OT1}{pzc}{m}{it}
\usepackage{mathtools}

\usepackage{stackengine}
\def\delequal{\mathrel{\ensurestackMath{\stackon[1pt]{=}{\scriptstyle\Delta}}}}

\usepackage{amsmath}

\usepackage{graphicx}
\usepackage[font={small}]{caption}
\usepackage{ptext}

\usepackage{algorithm}
\usepackage{algorithmic}
\usepackage{float}
\usepackage[utf8]{inputenc} 
\usepackage[T1]{fontenc}
\usepackage{url}
\usepackage{ifthen}
\usepackage{cite}

\usepackage{color}
\usepackage[dvipsnames]{xcolor}

\begin{document}
\title{\LARGE{Binary CEO Problem under Log-Loss with BSC Test-Channel Model}} 



\author{%
  \IEEEauthorblockN{\footnotesize{Mahdi Nangir}}
  \IEEEauthorblockA{\footnotesize{K.N. Toosi University of}\\
  	                \footnotesize{Technology, Tehran, Iran.} \\
                    \footnotesize{mahdinangir@ee.kntu.ac.ir}}
  \and             
  \IEEEauthorblockN{\footnotesize{Reza Asvadi}}
   \IEEEauthorblockA{\footnotesize{Shahid Beheshti University, }\\
                	\footnotesize{Tehran, Iran.} \\
                	\footnotesize{r\_asvadi@sbu.ac.ir}}
  \and
  \IEEEauthorblockN{\footnotesize{Mahmoud Ahmadian-Attari}}
  \IEEEauthorblockA{\footnotesize{K.N. Toosi University of} \\
  	\footnotesize{Technology, Tehran, Iran.} \\
  	\footnotesize{mahmoud@eetd.kntu.ac.ir}}
  \and
  \IEEEauthorblockN{\footnotesize{Jun Chen}}
  \IEEEauthorblockA{\footnotesize{McMaster University} \\ 
  	\footnotesize{Hamilton, ON, Canada.} \\
  	\footnotesize{junchen@mail.ece.mcmaster.ca}}
}


\maketitle

\begin{abstract}
In this paper, we propose an efficient coding scheme for the two-link binary Chief Executive Officer (CEO) problem under logarithmic loss criterion. The exact rate-distortion bound for a two-link binary CEO problem under the logarithmic loss has been obtained by Courtade and Weissman. We propose an encoding scheme based on compound LDGM-LDPC codes to achieve the theoretical bounds. In the proposed encoding, a binary quantizer using LDGM codes and a syndrome-coding employing LDPC codes are applied. An iterative joint decoding is also designed as a fusion center. The proposed CEO decoder is based on the sum-product algorithm and a soft estimator.
\end{abstract}

\section{Introduction}

The Chief Executive Officer (CEO) problem is defined by Berger \textit{et al.} for distributed source coding of multi-observations of a source corrupted by independent noises \cite{Berger96}. The CEO problem empirically emerges in wireless sensor networks, where a particular phenomenon is measured by some separate and independent sensors in a noisy environment. By using the compressed observations, a fusion center makes an estimation of the source at the receiver with an acceptable distortion between the original and the estimated source. In this paper, a soft reconstruction is considered.

In the last two decades, an abundant flurry of studies has been paid to address the theoretical bounds of the quadratic Gaussian CEO problem \cite{PTR04,oha12}. A tight upper bound on the sum-rate distortion function of the quadratic Gaussian CEO problem and the optimal rate allocation scheme are provided in \cite{CZB04}. Alternatively, studies like \cite{CB08} and \cite{BS09} present various coding schemes to achieve any point of the achievable rate-distortion region.

The case of a binary source with observations corrupted by the binary noises, called the binary CEO problem, has been paid less attention during these years. In general, its exact rate-distortion bound is an open problem in the information theory. The binary CEO problem appears in cooperative digital communication networks where some correlated remote sources are being sent to a central receiver via paralleled channels with independent noises. The most common criterion for measuring distortion in the binary case is the Hamming distortion measure. A lower bound for the achievable rate-distortion region of a two-link binary CEO problem is established in \cite{Tad} under the Hamming distortion. The Berger-Tung inner and outer bounds \cite{ELG}, are exploited for this case which are not tight under the Hamming distortion. The prior studies on the binary CEO in \cite{Tad} and \cite{RA14} consider that the correlated observations are transmitted through AWGN channels, and hence their encoders apply a channel coding to protect the transmitted data. Our goal is to achieve the maximum compression of the correlated noisy observations for sending through noiseless channels with minimum distortion.

Due to increasing demand for developing deep learning in upcoming complex networks, the logarithmic loss, or simply log-loss, has emerged as a useful criterion to measure distortion in many applications like machine learning, classification, and estimation theory. In this paper, we focus on the binary CEO problem under the log-loss criterion. This loss has been interpreted as the conditional entropy and the estimated symbols of the fusion center are soft. Moreover, it has been also shown that the log-loss is a universal criterion for measuring the performance of lossy source coding in \cite{AW15} and \cite{SRV17}.

Our main contributions in this paper can be considered in the context of coding and information theory. We assume a binary symmetric channel (BSC) as a test-channel for the lossy encoders in the binary CEO problem. Then, we obtain the optimal values of crossover probabilities of the test-channels for each BSC. Finally, an efficient coding scheme is proposed by utilizing the compound LDGM-LDPC codes and iterative message-passing algorithms. We show that the rate-distortion performance of the proposed coding scheme is close to the theoretical bounds. A full version of this paper including proofs and details are provided in \cite{arxiv18}.

The organization of this paper is as follows. In Section II, the system model and information theoretic aspect of the problem are provided. The designed encoding and decoding schemes are presented in Section III. Numerical results and discussions are given in Section IV. Finally, Section V draws the conclusion.

\section{Preliminaries}

Throughout this paper, the logarithm is to base $2$. Random variables, their realizations, and their alphabet set are depicted by the uppercase, lowercase, and calligraphic letters, respectively.

\subsection{System model}
\begin{figure}[t]
	\begin{center}
		\centering
		\includegraphics[width=2.7in,height=1.3in]{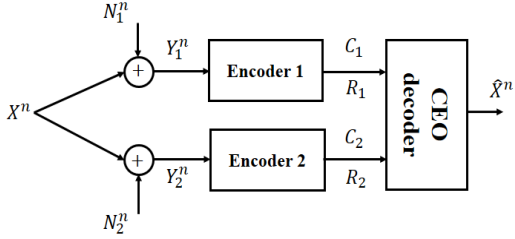}
	\end{center}
	\vspace{-10pt}
	\caption{\small{Block diagram of the two-link binary CEO problem.}}
	\label{CEO}
\end{figure}

Consider a communication system consisting of an independent and identically distributed (i.i.d.) binary symmetric source (BSS) and its two noisy observations being transmitted via two parallel links as depicted in Fig. \ref{CEO}. Let ${X}^n$, ${Y_1}^n$, and ${Y_2}^n$ denote a sequence of the BSS and two noisy observations of it, on the first and the second links, respectively. Observation noises ${N}_1^n$ and ${N}_2^n$ are independent from each other and are i.i.d. binary sequences generated by Bernoulli distributions with crossover parameters $p_1$ and $p_2$ associated to the first and the second links, respectively. Consider ${Y_1}^n$ and ${Y_2}^n$ are encoded to ${C_1}$ and ${C_2}$, and then they are sent to the CEO joint decoder. Note that ${C}_1 \leftrightarrow {Y}_1^n \leftrightarrow {X}^n \leftrightarrow {Y}_2^n \leftrightarrow {C}_2$ form a Markov chain. At the decoder, the sequence ${\hat X}^n$ is reconstructed in the joint CEO decoder by using $({C}_1,{C}_2)$.

The symbol-wise log-loss between a source symbol $x_j$ and its reconstruction $\hat{x}_j$ is defined as follows:
\begin{equation}
\label{eqlog}
d(x_j , \hat{x}_j) = \log \big({1 \over \hat{x}_j (x_j)}\big), \quad j=1,2,\cdots,n,
\end{equation}
where $\hat{x}_j(x_j)$ generally depends on $(c_1,c_2)$. The total value of log-loss between ${x}^n$ and $\hat {x}^n$ is obtained by uniform averaging over all symbols,
\begin{equation}
\label{eqlogg}
d({x}^n , \hat{x}^n) ={1 \over n} \sum_{j=1}^{n} \log \big({1 \over \hat{x}_j (x_j)}\big).
\end{equation}

\subsection{An Information Theory Perspective}
The rate-distortion theory objective implies to minimize the rate (distortion) subject to a fixed distortion (rate) value. By using the provided theoretical bounds in \cite{CW14}, the achievable rate-distortion region under the log-loss criterion for encoded sequences $U_1^n$ and $U_2^n$ is as follows:
\begin{align}
\label{sumrate-dist}
R_i& \ge I(Y_i;U_i|U_{i-3},Q), \ i=1,2, \\ \nonumber
R_1+R_2& \ge I({Y}_1,{Y}_2;{U}_1,{U}_2|Q),  \\ \nonumber
D &\ge H({X} | {U}_1,{U}_2,Q).
\end{align}
The cardinality bound implies that $|\mathcal{U}_i| \le |\mathcal{Y}_i|=2$ for $i=1,2$ and $|\mathcal{Q}| \le 4$. The dominant face of the achievable rate region is the sum-rate bound, hence the mentioned optimization problem of the rate-distortion theory is as follows:
\begin{align}
\label{opt1}
\underset{p(u_1|y_1,q)p(u_2|y_2,q)p(q)}{\text{min}}& \ \ I({U}_1,{U}_2;{Y}_1,{Y}_2 | Q), \\ \nonumber
\text{s.t.}& \ \  H({X}|{U}_1,{U}_2 , Q)= D_0,
\end{align}
where $H({X}|{Y}_1,{Y}_2) \le D_0 \le 1$. This optimization problem can be written in the following unconstrained form:
\begin{align}
\label{opt2}
\underset{p(u_1|y_1,q)p(u_2|y_2,q)p(q)}{\text{min}} H({X}|{U}_1,{U}_2 , Q) + {\mu} I({U}_1,{U}_2;{Y}_1,{Y}_2 | Q),
\end{align}
where $\mu$ is the Lagrangian multiplier. Note that 
\begin{align}
\label{opt21}
&H({X}|{U}_1,{U}_2 , Q) + {\mu} I({U}_1,{U}_2;{Y}_1,{Y}_2 | Q) \\ \nonumber
&=\sum_{q \in \mathcal{Q}} p(q) [H({X}|{U}_1,{U}_2 , Q=q) + {\mu} I({U}_1,{U}_2;{Y}_1,{Y}_2 | Q=q)] \\ \nonumber
& \ge \underset{q \in \mathcal{Q}}{\text{min}} \ \ H({X}|{U}_1,{U}_2 , Q=q) + {\mu} I({U}_1,{U}_2;{Y}_1,{Y}_2 | Q=q).
\end{align}
Therefore, for the purpose of characterizing the sum-rate-distortion function, there is no loss of generality in assuming that $Q$ is a constant, which eliminates $Q$ in (\ref{opt2}).

We shall assume that $p(u_i|y_i)$ is a BSC with crossover probability $d_i, \ i=1,2$, which is satisfied by the extensive numerical solution to (\ref{opt2}). Consequently, after some calculus manipulations, the rate and the distortion bounds are expressed by:
\begin{align}
\label{sumrate-dis}
R_i & \ge h_b(p*d)-h_b(d_i), \ i=1,2, \\ \nonumber
R \delequal R_1+R_2 & \ge 1+h_b(d*p)-h_b(d_1)-h_b(d_2), \\ \nonumber
D & \ge h_b(p_1*d_1)+h_b(p_2*d_2)-h_b(p*d),
\end{align}
where $d \delequal d_1*d_2=d_1(1-d_2)+d_2(1-d_1)$, $p \delequal p_1*p_2$, and $h_b(x)=-x \log_2x-(1-x) \log_2(1-x)$ is the binary entropy function.

\textbf{Theorem 1.} Bounds of distortion $D$ and sum-rate $R$ in (\ref{sumrate-dis}) are neither convex nor concave in terms of variables $(d_1,d_2)$.

\begin{figure}[t]
	\begin{center}
		\centering
		\includegraphics[width=2.7in,height=1.9in]{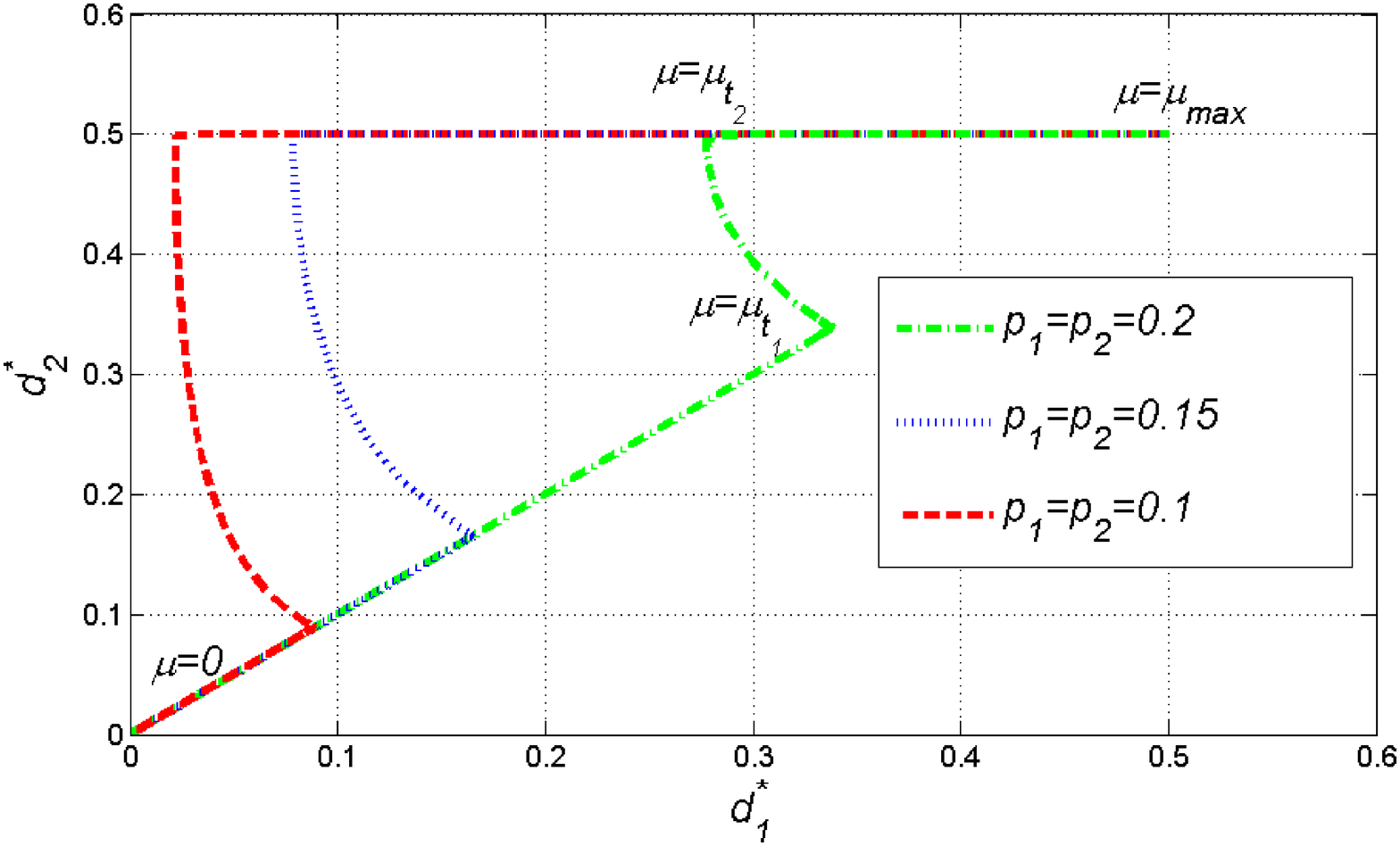}
	\end{center}
	\vspace{-10pt}
	\caption{\small{Location of the optimum points $(d_1^*,d_2^*)$.}}
	\label{fig1}
\end{figure}

Theorem $1$ implies that the optimization problem (\ref{opt1}) is not convex, even with considering BSC test-channel model for the encoders. Bounds in (\ref{sumrate-dis}) are two-variable functions of $(d_1,d_2)$, hence if we define the objective function $F(d_1,d_2) \delequal D+ \mu R$ for $\forall \mu \ge 0$, then for any fixed value of $\mu$ presume the minimum value of $F$ in the plane $[0,0.5] \times [0,0.5]$ that occurs in the point $(d_1^*,d_2^*)$. The location of these points is depicted in Fig. \ref{fig1} for some noise parameters. As it is seen in this figure, there are three different regions in each curve for any $(p_1,p_2)$. They comprise conditions as $d_1^*=d_2^*$ called Region $1$, $d_1^*<d_2^*$ called Region $2$, and $d_2^*=0.5$ called Region $3$. We implement our proposed coding scheme for some target distortion values $(d_1^*,d_2^*)$. An asymptotic analysis for the location of the optimum points $(d_1^*,d_2^*)$ is provided in \cite{arxiv18} at two extreme points $(d_1^*,d_2^*)=(0,0)$ and $(0.5,0.5)$. As a result, the variation range of the parameter $\mu$ has been obtained.

\textbf{Theorem 2.} The maximum value of $\mu$ occurs in $(R,D)=(0,1)$ when $(d_1^*,d_2^*)=(0.5,0.5)$ and it equals:

\begin{equation}
\label{mu}
\mu_{\max}=\max\{(1-2p_1)^2,(1-2p_2)^2\}.
\end{equation} 
Obviously, its minimum value equals $0$ and occurs in $(R,D)=(1+h_b(p),h_b(p_1)+h_b(p_2)-h_b(p))$ when $(d_1^*,d_2^*)=(0,0)$. If $(d_1^*,d_2^*) \to (0,0)$, then the location of optimum points follow from an explicit relation.

\textbf{Theorem 3.} For the optimization problem (\ref{opt2}) with BSC test-channel models, if $d_1 \rightarrow 0$ and $d_2 \rightarrow 0$, then:
\begin{equation}
\label{eq11}
d_2 \approx e^{K(K_2-K) \over (K_1-K)}d_1^{K_2-K \over K_1-K}.
\end{equation}

\textbf{Remark 1.} The slope of the tangent lines to the curve of location of optimum points in
$(d_1^*,d_2^*) = (0, 0)$ are, respectively, $1$, $\infty$, or $0$ when $p_1=p_2$ , $p_1<p_2$ , or $p_1>p_2$.

\section{The proposed coding scheme}
In this section, a practical coding scheme is proposed to achieve the rate-distortion bound of the binary CEO problem under BSC assumption for the encoders in each link. Actually, the encoders $1$ and $2$ are respectively modeled by BSCs with the crossover probabilities $d_1$ and $d_2$. In our proposed coding scheme, the Bias-Propagation (BiP) algorithm \cite{NAA17} is utilized for the binary quantization, and the Sum-Product (SP) algorithm and its modified version are employed for the syndrome-decoding.
\subsection{The Encoding Scheme}
A conventional rate-distortion quantizer can asymptotically achieve the compression rate $1-h_b(d_i)$ when distortion is assumed to be $d_i$, hence, it is impossible to get close to (\ref{sumrate-dis}) only by using a rate-distortion quantizer. Therefore, another lossless compressor should be employed after the quantizer for achieving the rate of (\ref{sumrate-dis}). We use an LDGM quantizer concatenated with a Syndrome-Generator (SG), inspired by the ``quantize-and-bin'' idea in the information theory.

For the two-link binary CEO problem, a binary Wyner-Ziv coding is considered in each link. Our proposed encoding includes two steps; first is mapping the observations ${Y}_1$ and ${Y}_2$ to the nearest codewords ${U}_1$ and ${U}_2$ in terms of the Hamming distance from two LDGM codebooks $1$ and $2$, respectively. This step is implemented by utilizing the BiP algorithm. In this step, the rate and distortion in $i$-th link are, respectively, denoted by $R_{i,1}={m_i \over n}$ and $d_{i,1}$, $i=1,2$. In the second step, syndromes of the codewords ${U}_i$ are generated by using the LDPC codes of rates $R_{i,2}={m_i - k_i \over n}, \ i=1,2$, and their non-zero entries are sent to the decoder as the lossy encoded sequences ${S}_1$ and ${S}_2$. These two steps are executed by using the compound LDGM-LDPC codes \cite{NAA17}. A compound LDGM-LDPC code in the $i$-th link includes nested LDGM and LDPC codes with the following parity-check matrices:
\begin{equation}
\label{matr}
H^{(i)}_{\text{LDPC}}=
\begin{bmatrix}
H^{(i)}_{\text{LDGM}} \\
\Delta H^{(i)} \\
\end{bmatrix},
\end{equation}
where $H^{(i)}_{\text{LDPC}}$ and $H^{(i)}_{\text{LDGM}}$ are, respectively, parity-check matrices of the LDPC and LDGM codes. Let assume their sizes are $(n-m_i+k_i) \times n$ and $(n-m_i) \times n$, respectively. Hence, the total rate of the $i$-th link is equal to $R_i=R_{i,1}-R_{i,2}={k_i \over n}$.

\subsection{The Joint CEO Decoder}
In the decoder, we propose a Joint Sum-Product (JSP) algorithm which is a modified version of the SP algorithm. In this algorithm, the received syndromes $s_1^{k_1}$ and $s_2^{k_2}$ are located in the check nodes of the LDPC codes. The JSP includes $r$ rounds and each round includes $l$ iterations. At the starting point of the JSP, initial LLRs in the variable nodes are set based on a random side information in each SP. At the end of each round, which includes update equations in the check and the variable nodes, the bit values are calculated according to the maximum likelihood (ML) decision rule of the SP algorithm. In the next round, these updated bit values in the variable nodes are used as a new side information for calculating successive initial LLR values. Finally, after $r$ rounds, $\hat u_1^n$ and $\hat u_2^n$ are decoded based on the ML decision rule of the SP algorithm in the variable nodes.

\begin{figure}[t]
	\centering
	\subfigure[{The proposed encoding scheme based on compound structure.}]{\label{ENC}
		\includegraphics[width=2.6in,height=1.5in]{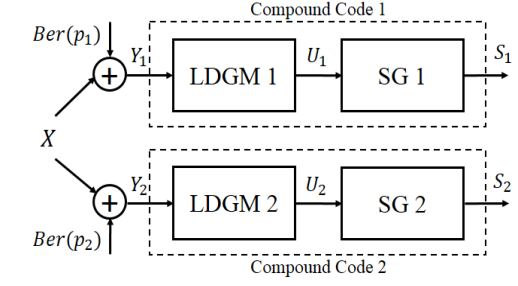}}
	\hspace{1mm}
	\subfigure[{Two-link joint decoding structure.}]{\label{jDEC}
		\includegraphics[width=2.2in,height=1.5in]{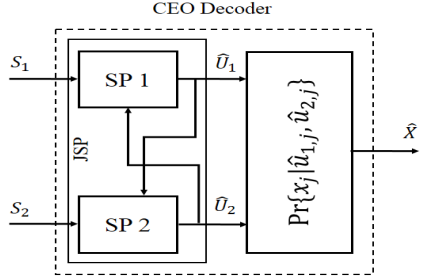}}
	\caption{The proposed coding scheme for achieving an intermediate point $(R_1^*,R_2^*)$.}
	\label{scheme}
\end{figure}

Consider the BER of the syndrome-decoding part is $d_{i,2}$, for $i=1,2$. Hence, the total distortion in each link equals $d_i=d_{i,1}*d_{i,2}$. An EXIT chart analysis is presented in \cite{KSA17} for a similar JSP decoder which shows the capacity approaching property with two parallel and collaborative SP decoders. The soft estimation $\hat x_j=\Pr\{x_j | \hat u_{1,j}, \hat u_{2,j}\}$ accomplishes the decoding process.
The block diagram of the proposed coding scheme is shown in Fig. \ref{scheme}. 

\subsection{A Practical Analysis for the Proposed Coding Scheme}
Some coding parameters are affected by the information theoretical limits, each of which should be considered in the code designing. In the following, any $\epsilon$ denotes a sufficiently small positive value. Assume that achieving the following intermediate point in the dominant face of the achievable rate region is desired,
\begin{equation}
\label{intermed}
(R_1^*,R_2^*)=\big(h_b(p*d)-h_b(d_1)+\delta,1-h_b(d_2)-\delta\big), 
\end{equation}
where $0 < \delta <1-h_b(p*d)$. In the proposed coding scheme for achieving (\ref{intermed}), the relation between the rate-distortion and the block lengths of each employed LDGM and LDPC codes are as follows:
\begin{align}
\label{eq19i}
R_{i,1}&={m_1 \over n}=1-h_b(d_{1,1})+\epsilon_{1,1}, \\ \nonumber 
R_{1,2}&={m_1-k_{1} \over n}=1-h_b(d_{1,1}*d_{2,1}*p_1*p_2)- \delta -\epsilon_{1,2}, \\ \nonumber
R_{2,1}&={m_2 \over n}=1-h_b(d_{2,1})+\epsilon_{2,1}, \\ \nonumber 
R_{2,2}&={m_2-k_{2} \over n}=\delta-\epsilon_{2,2},
\end{align}
which implies
\begin{align}
\label{eq21i}
R_1&=h_b(d_{1,1}*d_{2,1}*p_1*p_2)-h_b(d_{1,1})+ \delta +\underbrace{\epsilon_{1,1}+\epsilon_{1,2}}_{\epsilon_1} \\ \nonumber
&\stackrel{\text{(a)}}{\approx} h_b(d*p)-h_b(d_{1})+\delta+{\epsilon_1}, \\ \nonumber
R_2&=R_{2,1}-R_{2,2}={k_{2} \over n}=1-h_b(d_{2,1})-\delta+\underbrace{\epsilon_{2,1}+\epsilon_{2,2}}_{\epsilon_2} \\ \nonumber &\stackrel{\text{(a)}}{\approx} 1-h_b(d_{2})-\delta+\epsilon_{2},
\end{align}
where (a) follows from the continuity of the function $h_b(x)$. The above approximation expresses that achieving the theoretical rate bounds, for an intermediate point (\ref{intermed}), is possible by utilizing the proposed method. In the decoding side, from (\ref{eq19i}) and $0<\delta<1-h_b(p*d)$, we have:
\begin{align}
\label{rLD}
R_{1,2}&\approx 1-h_b(d*p)- \delta -\epsilon_{1,2}<1-h_b(d*p), \\ \nonumber
R_{2,2}&=\delta-\epsilon_{2,2}<1-h_b(d*p).
\end{align}
Therefore, the rates of LDPC codes in the SP$1$ and the SP$2$ algorithms are smaller than the capacity of the virtual channel between the side information $U_1$ and $U_2$. This implies that the SP algorithms can decode $U_1$ and $U_2$ with low BERs, i.e., $d_{i,2} \approx 0$ for $i=1,2$, for sufficiently large values of $n$, $r$, and $l$.

For the empirical distortion $D_{\text{em}}$ in (\ref{eqlogg}) with $\hat x_j(x_j)=\Pr\{x_j|u_{1,j},u_{2,j}\}$, we have
\begin{align}
\label{eq25}
D_{\text{em}}&={1 \over n} \sum_{j=1}^n \log [{1 \over \Pr\{x_j|u_{1,j},u_{2,j}\}}] \\ \nonumber 
&=\sum_{x,u_1,u_2} \Pr_{\text{em}}\{x,u_1,u_2\} \log [{1 \over \Pr\{x|u_1,u_2\}}],
\end{align}
where $\Pr_{\text{em}}\{x,u_1,u_2\}$ is the empirical distribution induced by $(x^n,u_1^n,u_2^n)$. The theoretical distortion bound is given by $D_{\text{th}}=H(X|U_1,U_2)$. Clearly, we have $D_{\text{em}} \approx D_{\text{th}}$ if $\Pr_{\text{em}}\{x,u_1,u_2\}$ gets close to $\Pr\{x,u_1,u_2\}$.

\section{Results and Discussions}

In this section, some numerical results are given for indicating the rate-distortion performance of the proposed coding scheme at different regions. For all of the LDPC codes, the optimized degree distributions over the BSC are employed \footnote{These degree distributions are available in \cite{SC07} for some rates.}. However, the check-regular and variable-Poisson LDGM codes nested with the LDPC codes are designed similar to the code design method used in \cite{NAA17}. In order to achieve some target optimum crossover probability pairs $(d_1^*,d_2^*)$, we have applied our proposed coding scheme with the lengths of $n=10^4, \ 10^5$ for the noise parameters $(p_1,p_2)=(0.15,0.15)$.

In the BiP algorithm, the parameters $t=0.8$, $\gamma_i \approx 2 R_{i,1}$ are selected for $i=1,2$. Maximum number of iterations in each round of this algorithm is set to be $25$. In a single SP algorithm, maximum number of iterations is set to be $100$. Also, in the JSP algorithm, $r=15$ and $l=40$. All of the reported values for the empirical distortions are averaged over $50$ runs. Parameters of the employed codes and their results are presented in Table \ref{t1}. The first and the second parts of this table are dedicated to the intermediate and the corner points of the sum-rate bound, respectively. The gap value is equal to difference between the empirical distortion $D_{\text{em}}$ and the theoretical distortion $D_{\text{th}}$ and it evaluates performance of the designed codes.

\begin{table*}[t]
	\caption{\footnotesize{NUMERICAL RESULTS OF THE PROPOSED ENCODING AND DECODING METHODS.}}
	\label{t1}
	\centering
	\vspace{-5pt}
	\begin{center}
		\scalebox{0.9}{
			\begin{tabular} {| c | c | c | c | c | c | c | c | c | c | c | c |}
				\hline
				$(p_1,p_2)$ &Region& $n$ & $m_1,m_2$ & $k_1,k_2$ & $d_{1,1},d_1$ & $d_{2,1},d_2$ &  $\mu$ & $R_{\text{th}}$ &  $D_{\text{th}}$ & $D_{\text{em}}$ &  $\text{Gap}$ \\
				\hline
				$\color{Green}(0.15,0.15)$ & $\color{Green}1$ & $\color{Green}10^4$ & $\color{Green}9200,9200$ & $\color{Green}8500,8500$& $\color{Green}0.0144,0.0175$ & $\color{Green}0.0144,0.0175$ & $\color{Green}0.168$& $\color{Green}1.6722$ & $\color{Green}0.4204$ & $\color{Green}0.4617$ & $\color{Green}0.0413$ \\
				
				$\color{Green}(0.15,0.15)$ & $\color{Green}1$ & $\color{Green}10^4$ & $\color{Green}5400,5400$ & $\color{Green}5100,5100$& $\color{Green}0.1028,0.1071$ & $\color{Green}0.1028,0.1071$ & $\color{Green}0.326$& $\color{Green}0.9898$ & $\color{Green}0.5925$ & $\color{Green}0.645$ & $\color{Green}0.0525$ \\
				
				$\color{Green}(0.15,0.15)$ & $\color{Green}2$ & $\color{Green}10^4$ & $\color{Green}5400,1300$ & $\color{Green}5300,1200$& $\color{Green}0.1028,0.1066$ & $\color{Green}0.3055,0.3078$ & $\color{Green}0.3854$& $\color{Green}0.6319$ & $\color{Green}0.7206$ & $\color{Green}0.7766$ & $\color{Green}0.056$ \\
				
				$\color{red}(0.15,0.15)$ & $\color{red}1$ & $\color{red}10^5$ & $\color{red}92000,92000$ & $\color{red}85000,85000$& $\color{red}0.012,0.015$ & $\color{red}0.012,0.015$ & $\color{red}0.168$& $\color{red}1.6722$ & $\color{red}0.4204$ & $\color{red}0.4451$ & $\color{red}0.0247$ \\
				
				$\color{red}(0.15,0.15)$ & $\color{red}1$ & $\color{red}10^5$ & $\color{red}54000,54000$ & $\color{red}51000,51000$& $\color{red}0.1003,0.1043$ & $\color{red}0.1003,0.1043$ & $\color{red}0.326$& $\color{red}0.9898$ & $\color{red}0.5925$ & $\color{red}0.6203$ & $\color{red}0.0278$ \\
				
				$\color{red}(0.15,0.15)$ & $\color{red}2$ & $\color{red}10^5$ & $\color{red}54000,13000$ & $\color{red}53000,12000$& $\color{red}0.1003,0.1037$ & $\color{red}0.3018,0.304$ & $\color{red}0.3854$& $\color{red}0.6319$ & $\color{red}0.7206$ & $\color{red}0.7494$ & $\color{red}0.0288$ \\
				\hline
				$\color{Green}(0.15,0.15)$ & $\color{Green}1$ & $\color{Green}10^4$ & $\color{Green}5400,5400$ & $\color{Green}4700,5400$& $\color{Green}0.1028,0.1076$ & $\color{Green}0.1028,0.1028$ & $\color{Green}0.326$& $\color{Green}0.9898$ & $\color{Green}0.5925$ & $\color{Green}0.6355$ & $\color{Green}0.043$ \\
				
				$\color{Green}(0.15,0.15)$ & $\color{Green}3$ & $\color{Green}10^4$ & $\color{Green}5400,-$ & $\color{Green}5400,-$& $\color{Green}0.1028,0.1028$ & $\color{Green}0.5,0.5$  & $\color{Green}0.4043$& $\color{Green}0.531$ & $\color{Green}0.7601$ & $\color{Green}0.7955$ & $\color{Green}0.0354$ \\
				
				$\color{Green}(0.15,0.15)$ & $\color{Green}3$ & $\color{Green}10^4$ & $\color{Green}1300,-$ & $\color{Green}1300,-$& $\color{Green}0.3055,0.3055$ & $\color{Green}0.5,0.5$& $\color{Green}0.4532$& $\color{Green}0.1187$ & $\color{Green}0.9427$ & $\color{Green}0.9835$ & $\color{Green}0.0408$ \\
				
				$\color{red}(0.15,0.15)$ & $\color{red}1$ & $\color{red}10^5$ & $\color{red}54000,54000$ & $\color{red}47000,54000$& $\color{red}0.1003,0.105$ & $\color{red}0.1003,0.1003$ & $\color{red}0.326$& $\color{red}0.9898$ & $\color{red}0.5925$ & $\color{red}0.6184$ & $\color{red}0.0259$ \\
				
				$\color{red}(0.15,0.15)$ & $\color{red}3$ & $\color{red}10^5$ & $\color{red}54000,-$ & $\color{red}54000,-$&  $\color{red}0.1003,0.1003$ & $\color{red}0.5,0.5$ & $\color{red}0.4043$& $\color{red}0.531$ & $\color{red}0.7601$ & $\color{red}0.7826$ & $\color{red}0.0225$ \\
				
				$\color{red}(0.15,0.15)$ & $\color{red}3$ & $\color{red}10^5$ & $\color{red}13000,-$ & $\color{red}13000,-$& $\color{red}0.3018,0.3018$ & $\color{red}0.5,0.5$ & $\color{red}0.4532$& $\color{red}0.1187$ & $\color{red}0.9427$ & $\color{red}0.9707$ & $\color{red}0.028$ \\
				\hline
		\end{tabular}}
	\end{center}
\end{table*}

\begin{figure}[t]
	\begin{center}
		\centering
		\includegraphics[width=3.4in,height=2.3in]{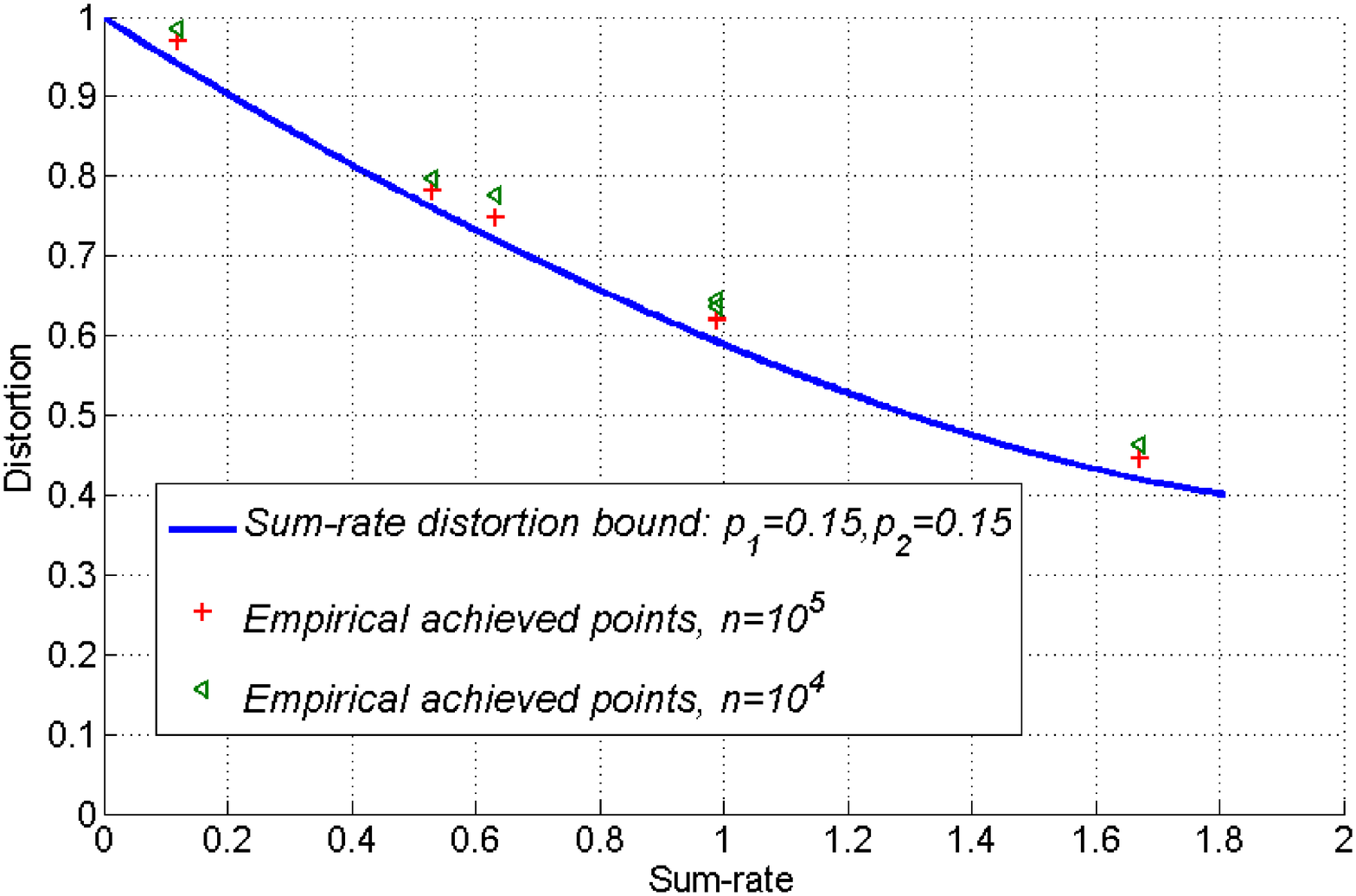}
	\end{center}
	\vspace{-5pt}
	\caption{\small{Performance of the sum-rate versus distortion for the proposed coding scheme.}}
	\label{RD1}
\end{figure}

For this case, the gap values is about $0.03$ to $0.06$ for the block length $10^4$, as indicated in Table. \ref{t1}. It is obvious that by increasing the target distortions, the gap values increase. As the block length is set to $10^5$, the gap value decreases in a range between $0.02$ to $0.03$. Performance of the sum-rate versus distortion is depicted in Fig. \ref{RD1} for the proposed coding scheme. The simulation results confirm that performance of the sum-rate in terms of distortion is very close to the theoretical bounds for the empirically achieved points. The theoretical bounds are asymptotically achievable by employing the proposed coding scheme as well.

\section{Conclusion}

In this paper, we investigated the two-link binary CEO problem under the log-loss, where the exact achievable rate-distortion region of the two-link binary CEO problem is known. By assuming the BSC test-channel model for the encoders, we found optimal values of the crossover probabilities. Next, we proposed a practical coding scheme based on the graph-based codes and message-passing algorithms. In the encoding side, a binary quantization and a syndrome-generation are utilized in each link for construction of the lossy compressed sequences. This was realized by the compound LDGM-LDPC codes. In the decoding, the SP algorithms based on the optimized LDPC codes for the BSCs were employed at the first step. Then, a soft decoder calculated the final reconstruction value in the form of a probability distribution. Our experimental simulation results confirmed that the proposed coding scheme asymptotically achieves the theoretical bound of the two-link binary CEO problem under the log-loss. For a finite block length, there remains a slight gap between the rate-distortion of the proposed scheme and the associated theoretical bound. 





\bibliographystyle{IEEEtran}

\bibliography{template_isit18}

\begin{thebibliography}{10}
\providecommand{\url}[1]{#1}
\csname url@samestyle\endcsname
\providecommand{\newblock}{\relax}
\providecommand{\bibinfo}[2]{#2}
\providecommand{\BIBentrySTDinterwordspacing}{\spaceskip=0pt\relax}
\providecommand{\BIBentryALTinterwordstretchfactor}{4}
\providecommand{\BIBentryALTinterwordspacing}{\spaceskip=\fontdimen2\font plus
\BIBentryALTinterwordstretchfactor\fontdimen3\font minus
  \fontdimen4\font\relax}
\providecommand{\BIBforeignlanguage}[2]{{%
\expandafter\ifx\csname l@#1\endcsname\relax
\typeout{** WARNING: IEEEtran.bst: No hyphenation pattern has been}%
\typeout{** loaded for the language `#1'. Using the pattern for}%
\typeout{** the default language instead.}%
\else
\language=\csname l@#1\endcsname
\fi
#2}}
\providecommand{\BIBdecl}{\relax}
\BIBdecl

\bibitem{Berger96}
T.~Berger, Z.~Zhang, and H.~Viswanathan, ``The {CEO} problem,'' \emph{IEEE
  Transactions on Information Theory}, vol.~42, no.~3, pp. 887--902, 1996.

\bibitem{PTR04}
V.~Prabhakaran, D.~Tse, and K.~Ramachandran, ``Rate region of the quadratic
  {G}aussian {CEO} problem,'' in \emph{IEEE International Symposium on
  Information Theory (ISIT), 2004.}\hskip 1em plus 0.5em minus 0.4em\relax
  IEEE, 2004, p. 119.

\bibitem{oha12}
Y.~Oohama, ``Distributed source coding of correlated {G}aussian remote
  sources,'' \emph{IEEE Transactions on Information Theory}, vol.~58, no.~8,
  pp. 5059--5085, 2012.

\bibitem{CZB04}
J.~Chen, X.~Zhang, T.~Berger, and S.~B. Wicker, ``An upper bound on the
  sum-rate distortion function and its corresponding rate allocation schemes
  for the {CEO} problem,'' \emph{IEEE Journal on Selected Areas in
  Communications}, vol.~22, no.~6, pp. 977--987, 2004.

\bibitem{CB08}
J.~Chen and T.~Berger, ``Successive {W}yner-{Z}iv coding scheme and its
  application to the quadratic {G}aussian {CEO} problem,'' \emph{IEEE
  Transactions on Information Theory}, vol.~54, no.~4, pp. 1586--1603, 2008.

\bibitem{BS09}
H.~Behroozi and M.~R. Soleymani, ``Optimal rate allocation in successively
  structured {G}aussian {CEO} problem,'' \emph{IEEE Transactions on Wireless
  Communications}, vol.~8, no.~2, pp. 627--632, 2009.

\bibitem{Tad}
X.~He, X.~Zhou, P.~Komulainen, M.~Juntti, and T.~Matsumoto, ``A lower bound
  analysis of {H}amming distortion for a binary {CEO} problem with joint
  source-channel coding,'' \emph{IEEE Transactions on Communications}, vol.~64,
  no.~1, pp. 343--353, 2016.

\bibitem{ELG}
A.~El~Gamal and Y.-H. Kim, \emph{Network information theory}.\hskip 1em plus
  0.5em minus 0.4em\relax Cambridge university press, 2011.

\bibitem{RA14}
A.~Razi and A.~Abedi, ``Convergence analysis of iterative decoding for binary
  {CEO} problem,'' \emph{IEEE Transactions on Wireless Communications},
  vol.~13, no.~5, pp. 2944--2954, 2014.

\bibitem{AW15}
A.~No and T.~Weissman, ``Universality of logarithmic loss in lossy
  compression,'' in \emph{IEEE International Symposium on Information Theory
  (ISIT), 2015}.\hskip 1em plus 0.5em minus 0.4em\relax IEEE, 2015, pp.
  2166--2170.

\bibitem{SRV17}
Y.~Shkel, M.~Raginsky, and S.~Verd{\'u}, ``Universal lossy compression under
  logarithmic loss,'' in \emph{IEEE International Symposium on Information
  Theory (ISIT), 2017}.\hskip 1em plus 0.5em minus 0.4em\relax IEEE, 2017, pp.
  1157--1161.

\bibitem{arxiv18}
M.~Nangir, R.~Asvadi, M.~Ahmadian-Attari, and J.~Chen, ``Analysis and code
  design for the binary ceo problem under logarithmic loss,'' \emph{arXiv
  preprint arXiv:1801.00435}, 2018.

\bibitem{CW14}
T.~A. Courtade and T.~Weissman, ``Multiterminal source coding under logarithmic
  loss,'' \emph{IEEE Transactions on Information Theory}, vol.~60, no.~1, pp.
  740--761, Jan 2014.

\bibitem{NAA17}
M.~Nangir, M.~Ahmadian, and R.~Asvadi, ``A binary {W}yner-{Z}iv code design
  based on compound {LDGM}-{LDPC} structures,'' \emph{IET Communications, DOI:
  10.1049/iet-com.2017.0032}, 2017.

\bibitem{KSA17}
M.~Khas, H.~Saeedi, and R.~Asvadi, ``{LDPC} code design for correlated sources
  using {EXIT} charts,'' in \emph{IEEE International Symposium on Information
  Theory (ISIT), 2017}.\hskip 1em plus 0.5em minus 0.4em\relax IEEE, 2017, pp.
  2945--2949.

\bibitem{SC07}
D.~H. Schonberg, \emph{Practical distributed source coding and its application
  to the compression of encrypted data}.\hskip 1em plus 0.5em minus 0.4em\relax
  University of California, Berkeley, 2007.

\end{thebibliography}

\end{document}